\pdfoutput=1
\documentclass[10pt, conference, letterpaper]{IEEEtran}
\IEEEoverridecommandlockouts
\usepackage{cite}
\usepackage{amsmath,amssymb,amsfonts}
\usepackage{multirow}
\usepackage[table]{xcolor}
\usepackage{colortbl,hhline}
\usepackage{graphicx}
\usepackage{textcomp}
\usepackage{xcolor}
\usepackage{algorithm}
\usepackage[noend]{algorithmic}
\usepackage{xspace}
\usepackage{url}
\usepackage{hyperref}
\usepackage{enumitem}

\def\BibTeX{{\rm B\kern-.05em{\sc i\kern-.025em b}\kern-.08em
    T\kern-.1667em\lower.7ex\hbox{E}\kern-.125emX}}
\newcommand{\sysname}{{\textsf{StarCast}}\xspace}

\begin{document}

\title{\sysname: A Secure and Spectrum-Efficient Group Communication Scheme for LEO Satellite Networks }

\author{\IEEEauthorblockN{Chaoyu Zhang,
Hexuan Yu,
Shanghao Shi,
Shaoyu Li,
Yi Shi,
Eric Burger,
Y. Thomas Hou, and 
Wenjing Lou}
\IEEEauthorblockA{Virginia Tech, VA, USA}
}

\maketitle

\begin{abstract}

Low Earth Orbit (LEO) satellite networks serve as a cornerstone infrastructure for providing ubiquitous connectivity in areas where terrestrial infrastructure is unavailable. With the emergence of Direct-to-Cell (DTC) satellites, these networks can provide direct access to mobile phones and IoT devices without relying on terrestrial base stations, leading to a surge in massive connectivity demands for the serving satellite. To address this issue, group communication is an effective paradigm that enables simultaneous content delivery to multiple users and thus optimizes bandwidth reuse. Although extensive research has been conducted to improve group communication performance, securing this communication without compromising its inherent spectrum efficiency remains a critical challenge. To address this, we introduce \sysname, a secure group encryption scheme for LEO satellite networks. Our solution leverages ciphertext-policy attribute-based encryption (CP-ABE) to implement fine-grained access control by embedding access policies directly within the ciphertext. Unlike standard secure communication approaches that require dedicated per-user channels and significantly deplete limited satellite spectrum resources, \sysname maintains efficient spectrum reuse within user groups while ensuring that only authorized users can access transmitted data. Additionally, it significantly reduces the costly key management overhead associated with conventional encryption schemes.

\end{abstract}

\begin{IEEEkeywords}
LEO Satellites, Spectrum Efficiency, Ciphertext-Policy Attribute-Based Encryption, Group Encryption
\end{IEEEkeywords}

\section{Introduction}\label{introduction}

LEO satellite networks have emerged as the key infrastructure for bridging connectivity gaps in remote and underserved regions. These networks support various applications, including remote access for mobile phones and IoT devices, disaster recovery efforts, and military communication. The integration of LEO satellite networks into the 3GPP ecosystem began with Release 17, which established global standards for NT networks, enabling satellite-based New Radio and IoT access for seamless service continuity~\cite{3GPPNTN}. With further enhancements planned in Release 18, LEO satellite networks are expected to play a central role in the evolution of 6G~\cite{heydarishahreza2024spectrumsharing}. The rise of DTC LEO satellites, such as Starlink V2, AST BlueWalker 3, and Lynk, has exponentially expanded user access~\cite{pasandi2024survey}. By enabling direct connectivity between satellites and devices, these systems eliminate the reliance on traditional ground infrastructure, extending connectivity to billions of users. This increasing demand underscores the urgent need to optimize spectrum efficiency to support the growing number of user connections while maintaining reliable, high-quality service.

However, LEO satellite communication systems face several critical constraints that limit their scalability in accommodating a massive number of user accesses. First, the onboard communication resources of LEO satellites are inherently limited due to restricted spectrum availability and the finite number of beams available for simultaneous connections, particularly in high-density areas~\cite{xiao2022leochallenge}. The emergence of DTC LEO satellites exacerbates the issue of limited licensed spectrum~\cite{yu2024pri}. Satellite network operators (SNOs) must share the already allocated 4G/5G bands with terrestrial mobile network operators (MNOs) under FCC/ITU regulations (Tab.~\ref{tab:SNO_MNO}). Additionally, the scarcity of orbital slots, which require approval from the ITU/FCC, imposes further constraints on the deployment of new satellites. Furthermore, the high capital investment required for developing, launching, and maintaining constellations presents a significant barrier to scaling these networks in a timely manner to meet sudden spikes in connectivity demand~\cite{curzi2020large, xie2021leo}.

\par \textbf{Motivation.} To address the critical challenge of sending data to multiple users, group communication plays a crucial role. Modern LEO satellites, such as Starlink~\cite{starlink_constellation}, increase their user access capacity by utilizing narrow spot beams, enabling them to serve multiple users simultaneously and reuse the same spectrum across different beams. Another technique to improve satellite spectrum efficiency is to use a broader beamwidth to enable multiuser coverage while using a single beam's spectrum to deliver the same content to many different users, such as multicast and broadcast communication. These approaches not only achieve high throughput, but also considerably increase an individual satellite's connectivity capacity. For example, the digital satellite television broadcast standard (DVB-S2X~\cite{DVB-S2X}) allows each spot beam to serve numerous customers simultaneously by broadcasting a single coded frame, thus optimizing resource allocation.

Multi-beam satellite systems have recently garnered considerable research interest, with extensive studies dedicated to improving satellite group communication performance~\cite{liu2020joint, yin2020rate, yan2020outage, zhu2018cooperative}. Despite these advancements, there remains a critical need to develop security solutions for LEO satellites in group communication scenarios while preserving the inherent spectrum efficiency. To protect information confidentiality, encryption is essential. If all receivers use the same secret key to decrypt the broadcasted ciphertext from the satellite, a single device compromise could lead to information leakage. Standard end-to-end encryption (E2EE) secures data and facilitates access control by employing a unique pair of cryptographic keys for each receiver~\cite{weiler2001secure, grosch2000framework}. However, this approach requires individual encrypted communication channels for each user, which hinders spectrum reuse within the group and undermines the efficiency expected from group communication. In a broadcast encryption system~\cite{barth2006privacy, fiat1994broadcast}, a ciphertext is protected by encrypting the group key using each group member's public key, thereby involving different keys for every user. However, scalability becomes a bottleneck as the number of system participants increases or as group membership changes dynamically. This significantly increases the number of required keys, complicating key distribution, management, and revocation.

\par \textbf{Our Solution.} To address these concerns, we propose \sysname, a secure and efficient group encryption scheme for LEO satellite networks that preserves the inherent spectrum efficiency of group communication. \sysname secures data and enables simultaneous transmission to all users covered by the same beam while ensuring that only legitimate users can decrypt the information.

\sysname utilizes CP-ABE~\cite{bethencourt2007ciphertext,waters2011ciphertext,chen2015improved,agrawal2017fame,riepel2022fabeo,meng2024fabesa,meng2024fease} as its cryptographic foundation, employing an attribute-based policy to facilitate scalable, fine-grained access control for a group of legitimate users. \sysname allows the server to encrypt data and specify an access policy for this information. The ciphertext, embedded with the access policy, is broadcast to various users within the beam but can only be decrypted by those whose attributes satisfy the access policy, eliminating the need for multiple individually encrypted channels for each user. \sysname significantly reduces the overhead associated with key management in large-scale systems, as users no longer require individual keys for each communication session. This attribute-based design enables our system to scale dynamically and efficiently. Furthermore, \sysname supports backward secrecy and revocability, and is resistant to collusion attacks, preventing adversaries from combining attribute keys from multiple non-members to access information that neither could access individually. \sysname ensures security while optimizing the utility of limited satellite spectrum resources by effectively combining multiple unicast communications with a physical-layer broadcast wireless signal for a group of users. Using this approach, the satellite can leverage phased array antennas to generate optimized and flexible beam directions and beamwidths tailored to group members, providing better signal strength compared to a large beam used for broadcasting to all users.

\par \textbf{Our Contribution:}

\begin{itemize}
    
    \item We propose \sysname, a secure and efficient group encryption scheme designed specifically for LEO satellite networks, addressing the critical need for enhanced connectivity capacity.
    
    \item \sysname leverages CP-ABE to provide fine-grained and scalable access control, significantly reducing the complexity of key management in dynamically changing and large-scale user systems.
    
    \item \sysname enhances both performance and efficiency in secure LEO satellite networks by enabling simultaneous ciphertext transmission to multiple users within the same beam and maximizing spectrum reuse.
    
    \item We prototype \sysname using the Starlink Standard Kit and conduct comparative evaluations against secure unicast and broadcast frameworks, demonstrating the spectrum efficiency of our design.
    
\end{itemize}

\section{Preliminary and Related Work}\label{system model}

\begin{figure}[t]
\centering
\includegraphics[width=0.9\linewidth]{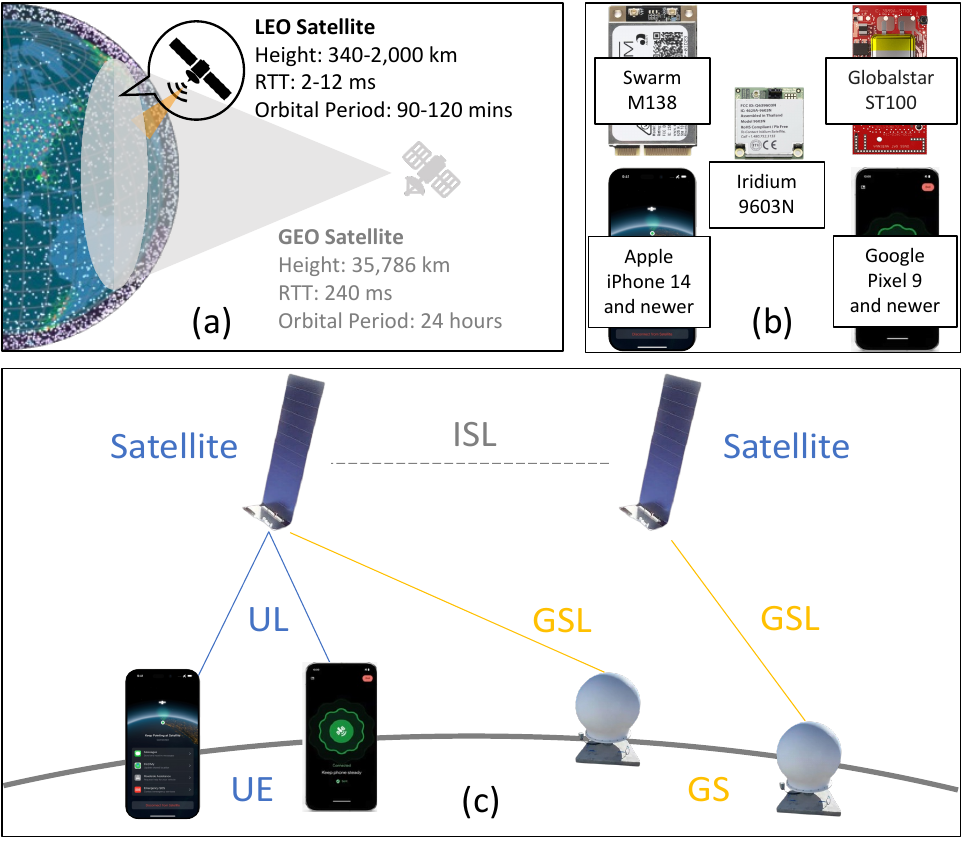}
\caption{(a) Starlink Coverage Map; (b) DTC Supported Mobile Phone and IoT Devices; (c) DTC LEO Satellite Network 
}
\label{starlink}
\vspace{-0.2cm}
\end{figure}

\begin{table}[t]
\centering
\caption{Comparison of Satellite Communication Providers and Specifications}
\begin{tabular}{ l|l|l|l}
\hline
\textbf{SNO/Partner}      & \textbf{\# of Satellites}   & \textbf{Spectrum/Band} & \textbf{Standard} \\ \hline
SpaceX/T-Mobile           & 2016         & MNO-band          & 3GPP      \\  
AST/AT\&T                 & 243          & MNO-band            & 3GPP      \\  
Kuiper/Verizon            & 3236         & Ka-band                & Proprietary        \\  
Globalstar/Apple          & 24           & L-band, S-band         & Proprietary        \\  
Iridium/Qualcomm          & 66           & L-band                 & Proprietary        \\ \hline
\end{tabular}
\label{tab:SNO_MNO}
\vspace{-0.2cm}
\end{table}

\subsection{LEO Satellite Constellation}

LEO satellite constellations orbit below 2000 km above the Earth~\cite{liou2008instability}, offering low-latency, high-speed connectivity compared to geostationary satellites (Fig.~\ref{starlink}(a)). One transformative feature of LEO satellites is their DTC capability, which enables mobile phones and IoT devices to connect directly to satellites without relying on terrestrial base stations. For instance, satellite network operators like Starlink V2 and other DTC-enabled systems seek to connect the roughly 2.6 billion people who do not have access to dependable internet~\cite{ITUunconnected} (Tab.~\ref{tab:SNO_MNO}). To achieve this, terminal devices that support the DTC satellite communication function (Fig.~\ref{starlink}(b)) can connect directly to an accessible LEO satellite via a User Link (UL). The satellite operates as an onboard radio access network (RAN) and transmits signals to a Ground Station (GS) via a GS-to-Satellite Link (GSL), which connects to the terrestrial core network (Fig.~\ref{starlink}(c)). Additionally, Starlink satellites utilize inter-satellite links (ISLs) powered by space laser communication, enabling direct satellite-to-satellite data transfer without the need for intermediate GS relays.

\subsection{Group Communication in Satellite Network}

The DTC capability of LEO satellites dramatically increases the number of users and devices accessing the network, leading to massive connectivity demands. Recent literature has focused on utilizing beams generated by phased array antennas to enhance spectrum efficiency, improve power density, and support higher data rates~\cite{kodheli2020satellite}. User grouping and beamwidth optimization have been proposed to address the trade-offs between serving more users simultaneously and maintaining adequate signal strength~\cite{yin2020rate, liu2020joint} for downlink multicast communications. Various beamforming and pre-coding strategies have also been explored, including optimization algorithms that minimize power consumption while maintaining the quality of service~\cite{zou2021robust, van2022user, wang2021resource}.

\subsection{Secure Group Communication}

Despite extensive research to enhance group communication efficiency, the challenge of ensuring secure communication without compromising spectrum efficiency remains. As next-generation communication systems evolve, the security and privacy of upcoming 6G networks are increasingly critical~\cite{du2023ucblocker, yuaaka, du2022mobile, li2023bijack, zhang2023mindfl, zhang2024hermes, zhang2025IDSSurvey, Yu2025closing, zhang2025sentinel, zhang2024state}. Secure group communication can be achieved through centralized key management and end-to-end session encryption~\cite{tedeschi2022satellite, liu2022secure, godhwani2011use}. Another approach, broadcast encryption, ensures communication security and access control by encrypting the group key using each member’s public key. Various broadcast encryption schemes embed these encrypted keys in the ciphertext~\cite{weiler2001secure, grosch2000framework, barth2006privacy, fiat1994broadcast}. Additionally, several efforts focus on improving protection scheme efficiency~\cite{yu2021fpga, zhang2021high, fan2019gpu}. Attribute-based encryption originated as a fuzzy variant of identity-based encryption~\cite{sahai2005fuzzy}. In CP-ABE schemes~\cite{bethencourt2007ciphertext, waters2011ciphertext, chen2015improved, agrawal2017fame, riepel2022fabeo, meng2024fabesa, meng2024fease}, a user's private key is associated with specific attributes, and ciphertexts are encrypted according to an access policy. Decryption is only possible if the user's attributes meet the policy requirements.

\subsection{Cryptographic Preliminaries}

\subsubsection{Bilinear map or pairing $e$}

The cryptographic design of \sysname is based on CP-ABE. This scheme leverages bilinear maps as the mathematical foundation to enable efficient cryptographic operations, fine-grained access control, and provable security. For all $u, v \in \mathbb{Z}$, $x \in \mathbb{G}_1$, and $y \in \mathbb{G}_2$, it holds that $e(x^u, y^v) = e(x, y)^{uv}$. 
Besides, $e$ is non-degenerate, i.e., $e(x, y) \neq 1$. We define $\textbf{Gen}$ as an asymmetric pairing group generator that, on input $1^{\lambda}$, outputs three groups $\mathbb{G}_1$, $\mathbb{G}_2$, and $\mathbb{G}_T$ of prime order $p = \Theta(\lambda)$. These groups come with a non-degenerate and efficiently computable bilinear map $e: \mathbb{G}_1 \times \mathbb{G}_2 \rightarrow \mathbb{G}_T$.
$g$ and $h$ are the group generators for $\mathbb{G}_1$ and $\mathbb{G}_2$, respectively. The CP-ABE scheme utilized in \sysname is based on Type-III pairings (i.e., asymmetrical prime order pairings).

\subsubsection{Access Policy and Attributes}

An access policy $\mathcal{A}$ specifies how attributes are managed to control access to the encrypted secret~\cite{riepel2022fabeo}, and it is typically embodied in a policy that specifies which combinations of user attributes are required to decrypt a ciphertext. Access policy $\mathcal{A}$ is a collection of non-empty subsets of $\mathcal{U}$, $\mathcal{A} \subseteq 2^\mathcal{U} \setminus \{0\}$, where $\mathcal{U}$ denotes the universe of attributes. $\mathcal{A}$ can be effectively modeled using Boolean formulas involving AND and OR gates. If all inputs corresponding to attributes in $S$ are set to true and all other inputs are set to false, then a set of attributes $S \subseteq \mathcal{U}$ meets a formula.

A generalization of the Boolean formula can be achieved by monotone span programs (MSPs), also known as a liner secret sharing scheme~\cite{lewko2010fully, rouselakis2013practical}. The \textit{monotonicity} is defined as for all $\mathcal{P}, \mathcal{Q} \subseteq \mathcal{U}$ that $\mathcal{P} \subseteq \mathcal{Q}$, $\mathcal{P} \in \mathcal{U}$ implies $\mathcal{Q} \in \mathcal{U}$.  \textit{Monotonicity} implies that if an authorized user has additional attributes, their access remains valid and their privilege remains true.

An MSP is given by a matrix $\mathbf{M}$ of size $n_1 \times n_2$ over $\mathbb{Z}_p$ and a mapping $\pi : \{1, \ldots, n_1\} \rightarrow \mathcal{U}$.  A monotone boolean formula $F$ can be converted into an MSP $(\mathbf{M}, \pi)$~\cite{lewko2011unbounded}. Every element in the matrix $\mathbf{M}$ is either 0, 1, or -1. Each row of matrix $\mathbf{M}$ represents an input in $F$, and the number of columns is equal to the number of AND gates in $F$. 

Suppose $S$ is a set of attributes, and $I = \{i \mid i \in \{1, \ldots, n_1\}, \pi(i) \in S\}$ is the set of rows in $\mathbf{M}$ that belong to $S$. 
We define the acceptance of $S$ by $(\mathbf{M}, \pi)$ as the existence of a linear combination of rows in $I$ that yields the vector $(1, 0, \ldots, 0)$ such that
$\sum_{i \in I} \gamma_i (\mathbf{M})_i = (1, 0, \ldots, 0)$, where $(\mathbf{M})_i$ denotes the $i$th row of $\mathbf{M}$.

\section{System Model and Threat Model}

\subsection{System Model}

\begin{figure}[t]
\centering
\includegraphics[width= 0.9\linewidth]{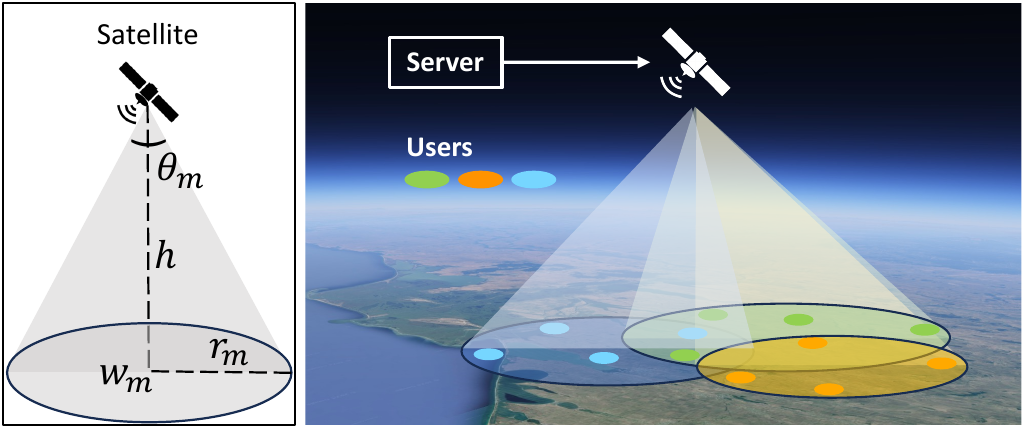}
\caption{\sysname system model}
\label{fig:p1}
\vspace{-0.2cm}
\end{figure}

We consider a scenario for downlink group communication through LEO satellites (Fig.~\ref{fig:p1}, right). The system comprises three key components: the server, $N$ ground users, and the LEO satellite. The server utilizes a terrestrial ground station to access the on-orbit satellite, which transmits encrypted data to different groups of users. The server determines the access policy for users based on their corresponding privileges, and all users possess secret keys generated by the server. In our system, we assume that the LEO satellite can obtain the location information of all users and is aware of their access privileges as provided by the server. The satellite is equipped with advanced phased array technology at a height $h$, enabling it to generate beams with flexible directionality. The coordinates of the center of a beam are represented by $\mathbf{w}_m$, while the beam's coverage radius is denoted by $r_m$. The adjustable beamwidth, $\theta_m$, expressed in degrees, is adaptive to provide optimal signal strength for a legitimate group of users (Fig.~\ref{fig:p1}, left). The server employs group communication to send encrypted data to users, ensuring that only legitimate recipients within the beam can decrypt it. The system is capable of managing the mobility of user devices and handling beam handovers when devices are covered by multiple beams, ensuring a smooth transition of the wireless signal from one beam to another. LEO satellites, equipped with phased array antennas, maintain continuous communication sessions with user devices, even during ground station handovers between satellites.

\subsection{Threat Model}

We assume that the satellite communication service is controlled and provided by an honest-but-curious third party, such as Starlink. Consequently, a security mechanism is required to preserve data confidentiality and achieve access control. We assume that an attacker can intercept wireless signals transmitted between the satellite and user devices, potentially gaining access to ciphertext. Our primary concern is the encryption of the plaintext data and ensuring access control, so only the legitimate group of users can decrypt the ciphertext and access the original information. Other types of attacks, such as jamming or physical destruction of user devices, are beyond the scope of this research and require separate studies to address.

\section{Design of \sysname}\label{Design Details}

\begin{figure*}[t]
\centering
\includegraphics[width=0.85\linewidth]{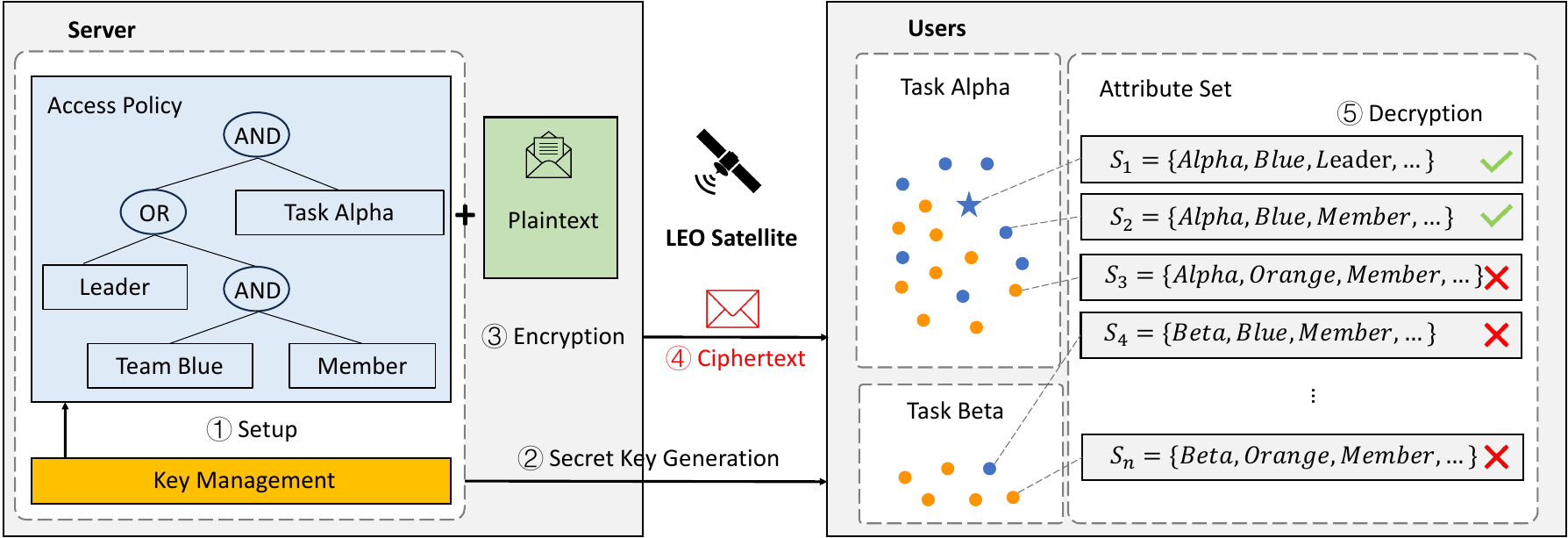}
\caption{\textbf{System Design:} In our system, the server defines an access policy, and only those users whose attributes meet it can decrypt the data. An LEO satellite utilizes group communication to send the ciphertext to the users of the legitimate group simultaneously.}
\label{fig:p2}
\vspace{-0.5cm}
\end{figure*}

\subsection{Scheme Overview}

The key idea of \sysname is to secure the group communication scheme for LEO satellite networks while preserving the inherent spectrum efficiency of group communication. At the physical layer, \sysname broadcasts encrypted data to all users covered by the same beam simultaneously. At the presentation layer, the cryptographic design ensures data confidentiality and guarantees that only legitimate users can decrypt the information. To achieve this, the server in \sysname sends the ciphertext $\mathit{c}$ for group communication, attaching it with an access policy $\mathcal{A}$. On the user side, a secret key $\mathit{sk}$ associated with a set of attributes can decrypt the ciphertext only if the set of attributes $S$ satisfies the access policy $\mathcal{A}$. Our formal definitions of the cryptographic design follow~\cite{riepel2022fabeo, agrawal2017fame}, whose security properties and proofs are well established.

The high-level concept of the \sysname cryptographic scheme with MSPs is described as follows. Each secret key is associated with the user's attribute set, $S$. For each attribute $x \in S$, there is a corresponding component $sk_y$ in the key. These components are composed of elements from group $\mathbb{G}_2$. To prevent users from colluding multiple keys to decrypt a ciphertext that none of them could individually decrypt, the policies of these components and their interconnections must be tied properly. Meanwhile, each ciphertext contains components from a group $\mathbb{G}_1$. For a given access policy represented by an MSP, each row of the MSP matrix corresponds to a component $c_i$, where $i$ is the $i$-th row of the matrix $\mathbf{M}$. This part can mask the $i$-th row's share by some specific numbers, which have to correspond to $sk_{\pi(i)}$, by the form of the component of the key. Thus, it allows a user with attribute $\pi(i)$ to reveal the $i$-th share during $\textbf{Dec()}$. This ensures that only users with a specific set of policy-compliant attributes can decrypt the information while also preventing any type of collusion attack. The public parameters, established during $\textbf{Setup()}$, provide the necessary values for both the ciphertext and key components. It is crucial to assign distinct group elements to each attribute to prevent any single key component from revealing multiple shares in a ciphertext.

\subsection{Workflow} \label{ABE example}

We illustrate the workflow of \sysname in Fig.~\ref{fig:p2}. For initialization, the server first sets up essential security parameters and generates a public key and a master key at step \textcircled{1}. Then, it generates and distributes the corresponding secret keys for each user based on their attributes at step \textcircled{2}. During group communication, the server encrypts data at step \textcircled{3} according to the access policy, which is formulated by the data management rules and users' privilege. At step \textcircled{4}, the server uploads ciphertexts to the LEO satellites, which serve as critical infrastructure for utilizing group communication to deliver these messages to users. The LEO satellites, equipped with advanced phased array technology, can generate the beam with flexible directionality and adjustable beamwidth, leveraging vendor-provided algorithms to ensure optimal signal strength for the legitimate user group. Finally, at step \textcircled{5}, the users decrypt the ciphertext data if and only if their attributes satisfy the corresponding access policies.

For example, we consider the server to define the access policy for users as follows: Only the team leaders or Team Blue's members participating in task Alpha can decrypt the message, i.e., $\mathcal{A}=(\text{"Alpha"} \text{AND} (\text{"Leader"} \text{OR} (\text{"Blue"} \text{AND} \text{"Member"})))$. As a result, the user with an attribute set $S_1 = \{\text{Alpha}, \text{Blue}, \\ \text{Leader}, \ldots\}$ can decrypt the message, but another user with an attribute set  $S_4 = \{\text{Beta}, \text{Blue}, \text{Member}, \ldots\}$ cannot.

\subsection{Protocol Details}

\par \noindent\underline{\textbf{Setup}($1^\lambda$)}: On inputting a security parameter $\lambda$, the server generates $(p, \mathbb{G}_1, \mathbb{G}_2, \mathbb{G}_T, e, g, h)$. $\mathbb{G}_1, \mathbb{G}_2, \mathbb{G}_T$ are three groups with prime order $p$, while $g$ and $h$ are two generators for $\mathbb{G}_1$ and $\mathbb{G}_2$, respectively.  
Server picks $\alpha_1, \beta_1, \alpha_2, \beta_2 \leftarrow_R \mathbb{Z}_p^*$ and $\delta_1, \delta_2, \delta_3 \leftarrow_R \mathbb{Z}_p$, computes the public key $\mathit{pk}$ and the master key $\mathit{mk}$:
\begin{align}
    pk: &(h, T_1 := e(g, h)^{\delta_1 \alpha_1 + \delta_3}, \\ 
    & H_1 := h^{\alpha_1}, T_2 := e(g, h)^{\delta_2 \alpha_2 + \delta_3}, H_2 := h^{\alpha_2})\nonumber\\
    mk: &(\alpha_1, \beta_1, \alpha_2, \beta_2, g, h, g^{\delta_1}, g^{\delta_2}, g^{\delta_3})
\end{align} 

\noindent \underline{\textbf{KeyGen}($\textit{mk}, S$)}: The server processes the attribute sets for each user and generates their respective secret keys. It picks $\rho_1, \rho_2 \leftarrow_R\mathbb{Z}_p$ and uses $h, \beta_1, \beta_2$ from $\textit{mk}$ to compute:
\begin{align}
        \textit{sk}_0 := (h^{\beta_1 \rho_1}, h^{\beta_2 \rho_2}, h^{\rho_1 + \rho_2})
\end{align}

Arbitrary binary strings can be mapped to elements of the group $\mathbb{G}_1$ using a hash function $\mathcal{H}$. $\mathcal{H}$ receives two kinds of input: strings and numbers. It takes the form of 
 $(\xi,\ell,t)$ or $(\rho,\ell,t)$, where $\xi$ is an arbitrary string, $\rho$ is a positive integer, $\ell \in \{1, 2, 3\}$ and $t \in \{1, 2\}$. 
 These two inputs are denoted, for simplicity, as $\xi\ell t$ and $0\rho\ell t$, respectively. For easy differentiation between them, append $0$ at the beginning of the second element. For all $x \in S$ and $t = 1, 2$, \noindent where $\sigma_x \leftarrow_{R}\mathbb{Z}_p$, compute:
\begin{align}
    \textit{sk}_{x,t} := \mathcal{H}(x1t)^{\frac{\beta_1 \rho_1}{a_t}} \cdot \mathcal{H}(x2t)^{\frac{\beta_2 \rho_2}{a_t}} \cdot \mathcal{H}(x3t)^{\frac{\rho_1 + \rho_2}{a_t}} \cdot g^{\frac{\sigma_x}{a_t}}.
\end{align}
 
\noindent Set $\textit{sk}_x := (\textit{sk}_{x,1}, \textit{sk}_{x,2}, g^{-\sigma_x})$, again for $t = 1, 2$, where $\sigma' \leftarrow_R \mathbb{Z}_p$, compute:

\begin{align}
        \textit{sk}^{'}_{t} := g^{d_t} \cdot \mathcal{H}(011t)^{\frac{\beta_1 \rho_1}{a_t}} \cdot \mathcal{H}(012t)^{\frac{\beta_2 \rho_2}{a_t}} \cdot \mathcal{H}(013t)^{\frac{\rho_1 + \rho_2}{a_t}} \cdot g^{\frac{\sigma^{'}}{a_t}}.
\end{align}

\noindent Set $\textit{sk}' = (\textit{sk}'_1, \textit{sk}'_2, g^{\delta_3}, g^{-\sigma'})$, 
server generates the secret key for a user as $(\textit{sk}_0, \{\textit{sk}_x\}_{x \in S}, \textit{sk}')$.
\par \noindent \underline{\textbf{Enc}($\textit{pk}, (\mathbf{M}, \pi), m$)}: server picks $s_1, s_2 \leftarrow_R\mathbb{Z}_p$ and generates:
\begin{align}
    c_0 := (H_1^{s_1}, H_2^{s_2}, h^{s_1+s_2})
\end{align}

When the secret message $m$ is split into shares using the MSP matrix $\mathbf{M}$, assuming $\mathbf{M}$ has $n_1$ rows and $n_2$ columns, such that  $(\mathbf{M})_{i,j}$ denotes the $(i,j)$-th element of $\mathbf{M}$. Then, for $i = 1, \ldots, n_1$ and $\ell = 1, 2, 3$, calculate:

\begin{align}
     c_{i,\ell} := \mathcal{H}(\pi(i) \ell 1)^{s_1}  \mathcal{H}(\pi(i) \ell 2)^{s_2} \cdot  \nonumber\\
      \prod_{j=1}^{n_2} [\mathcal{H}(0j \ell 1)^{s_1} \cdot 
      \mathcal{H}(0j\ell 2)^{s_2} ]^{(\mathbf{M})_{i,j}},      
\end{align}

\noindent and set $c_i := (c_{i,1}, c_{i,2}, c_{i,3})$, $c' := T_1^{s_1} \cdot T_2^{s_2} \cdot m$. The ciphertext is $(c_0, c_1, \ldots, c_{n_1}, c')$.

\par \noindent \underline{\textbf{Dec}($\textit{pk}, c, \textit{sk}$)}: if the set of attributes $S$ in $\textit{sk}$ satisfies the MSP $(\mathbf{M}, \pi)$ in $c$, then there exist constants (i.e., coefficients) $\{ \gamma_i \}_{i \in I}$ that satisfy equation $\sum_{i \in I} \gamma_i (\mathbf{M})_i = (1, 0, \ldots, 0)\label{eq1}$. A user can decrypt the message $m$ by computing $z_1/z_2$ ($\textit{sk}_{0,1}$, $\textit{sk}_{0,2}$, $\textit{sk}_{0,3}$ denote the first, second, and third elements of $\textit{sk}_0$, which is same for $c_0$):  
\begin{align}
    z_1 :=   &{c}' \cdot e \left( \prod_{i \in I} c^{\gamma_i}_{i,1}, \textit{sk}_{0,1} \right)
    \cdot e \left( \prod_{i \in I} c^{\gamma_i}_{i,2}, \textit{sk}_{0,2} \right) \nonumber \\
    &\cdot 
    e\left( \prod_{i \in I} c^{\gamma_i}_{i,3}, \textit{sk}_{0,3} \right),
\end{align}
\begin{align}
    z_2 := &e \left( \textit{sk}'_1, \prod_{i \in I} sk^{\gamma_i}_{\pi(i),1}, c_{0,1} \right) 
    \cdot e \left( \textit{sk}'_2, \prod_{i \in I} sk^{\gamma_i}_{\pi(i), 2}, c_{0,2} \right) \nonumber \\
    &\cdot 
    e \left( \textit{sk}'_3, \prod_{i \in I} sk^{\gamma_i}_{\pi(i),3}, c_{0,3} \right).
\end{align}

The plaintext $m$ sent to the encryption process is an element of the target group. 
It is possible that, in reality, the data is too large to fit into a single $\mathbb{G}_T$ element, and it would be quite costly to disassemble the message and encrypt it using attribute-based encryption.
To efficiently process large messages, one effective solution is to utilize the hash result of $T_1^{s_1}T_2^{s_2}$ as a symmetric session key. This key can then be utilized to efficiently encrypt the message using an ordinary symmetric encryption scheme, e.g., AES. Therefore, the computational cost of encrypting data of any size using this CP-ABE approach can be reduced to the expense of one single ABE encryption process.

\subsection{Scheme Discussion}

\sysname can support \textit{backward secrecy} and \textit{revocability}, ensuring that any new user cannot decrypt the ciphertext before joining the group. To accomplish this, the master key $\mathit{mk}$ is updated before each new user joins.
Additionally, the secret keys $\mathit{sk}$ for all current legitimate users are regenerated. Member revocation is realized in a similar manner by updating the $\mathit{mk}$ for all users, excluding those that need to be revoked.

\textbf{Handling Dynamically Changing User Groups.} \sysname is well-suited for the dynamic nature of LEO satellite communication, where the intended recipients (i.e., policies) of different messages may change dynamically.  The user’s $\mathit{sk}$ is valid for decrypting any ciphertext as long as the user’s attributes satisfy the policy of that specific ciphertext. users do not need separate keys for each ciphertext; instead, they use the same attribute-based key across different policies, decrypting only those ciphertexts for which their attributes satisfy the policy. As long as a user’s attributes have not changed, there is no need to update its $\mathit{sk}$, their secret key remains the same, regardless of how many different policies are used to encrypt different messages (for example, the access policy for a new encrypted data may become more restrictive or inclusive). 
The ability to decrypt one ciphertext does not affect the ability to decrypt another. Each ciphertext is evaluated independently based on its policy and the recipient’s attributes.

\section{Evaluation}

\begin{figure}[t]
\centering
\includegraphics[width=\linewidth]{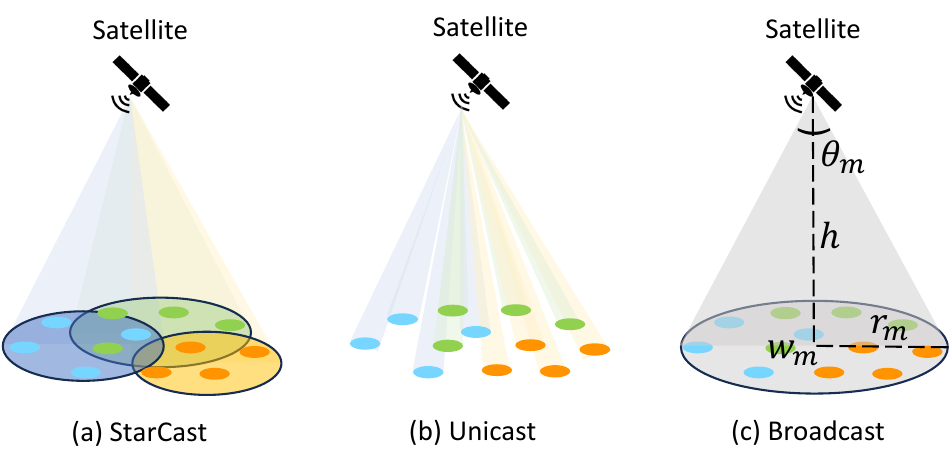}
\caption{Comparison of \sysname, Unicast, and Broadcast Schemes for Secure Communication in LEO Satellite Networks}
\label{fig: casts}
\vspace{-0.2cm}
\end{figure}

To evaluate the spectrum efficiency of our design, we compare the total data rate for legitimate users (those who meet the decryption policy for a specific ciphertext) across different secure LEO satellite communication methods. We prototype and evaluate the latency of \sysname using the Starlink LEO satellite network, and analyze the computational cost of the cryptographic scheme. Additionally, we compare the cost across conventional unicast encryption, broadcast encryption, and \sysname for secure LEO satellite group communications.

\subsection{Total Data Rate from Legitimate Users}

We consider the secure satellite communication schemes illustrated in Fig.~\ref{fig: casts} and follow our system model assumption. These users are grouped according to various information access privileges (as indicated by different colors). In a unicast framework (Fig.~\ref{fig: casts}(b)), the satellite directs a narrow, focused beam to a single user. The broadcast framework (Fig.~\ref{fig: casts}(c)) involves the satellite generating a wide beam that covers the entire geographical area. \sysname strikes a balance between the two (Fig.~\ref{fig: casts}(a)), where the satellite generates beams of intermediate width, targeting specific groups of users with shared content.

To analyze the sum of legitimate users' data rates, we follow the performance evaluation model from~\cite{liu2020joint}. We assume that the same group of users utilizes the line-of-sight path to the satellite and requests the same content. Although satellite communication can support multiple beams simultaneously, we simplify the analysis by assuming that only one beam is used for transmission at any given time. The satellite provides communication services to a total of $N$ users. They are grouped into $M$ groups, and each group is served sequentially using time division. Let $\mathcal{G}_m$ represent the set of users in group $m$, where $m \in {1, 2, \dots, M}$, and the number of users in each group is given by $|\mathcal{G}_m|$, such that $\sum_{m=1}^{M} |\mathcal{G}_m| = N$. The number of groups $M$ is determined by the information access privileges based on the attribute sets of the users. A beam sample is shown in Fig.~\ref{fig: casts}(c), where $\mathbf{w}_m$ represents the coordinates of the center of the beam serving group $m$. The adjustable coverage radius of the beam for group $m$, denoted by $r_m$. Here, $h$ represents the height of the satellite, and $\theta_m$ is the beam angle, expressed in degrees, which measures the direction in which the gain falls to half of its maximum. The beam coverage radius $r_m$ is related to the beam angle $\theta_m$ by $ r_m = h \tan\left(\frac{\theta_m}{2}\right).$

The data rate $C_i$ for user $i$ is determined by the signal-to-noise ratio (SNR). The received power $P_r^i$ can be calculated using the Friis transmission equation:

\begin{equation}
    P_i^r = \frac{1}{2} \alpha G_i^t G^r P^t \left(\frac{\lambda}{4\pi h}\right)^2,
    \label{Friis transmission Eq}
\end{equation}
 
where $\alpha$ is the power attenuation factor, $G_i^t$ and $G^r$ are the gains of the transmitter and receiver antennas, respectively, $P^t$ is the transmitted power of the satellite, and $\lambda$ is the wavelength. The transmitter antenna gain $G_t^i$ is a function of the beamwidth $\theta_i$ and can be approximated as $ G_i^t = \eta \left(\frac{70\pi}{\theta_i}\right)^2$, where $\eta$ is the antenna efficiency, and $\theta_i$ is the beam angle. Suppose group $\mathcal{G}_m$ is the legitimate group, the sum of users' data rate can be expressed as:

\begin{equation}
C_m = \sum^{|\mathcal{G}_m|}_{i=1}  B \log_2 (1 + \frac{P_i^r}{N_0 B}), 
\end{equation}

where $B$ is the channel bandwidth, and $N_0$ is the noise power spectral density.

\begin{figure}[t]
\centering
\includegraphics[width=0.85\linewidth]{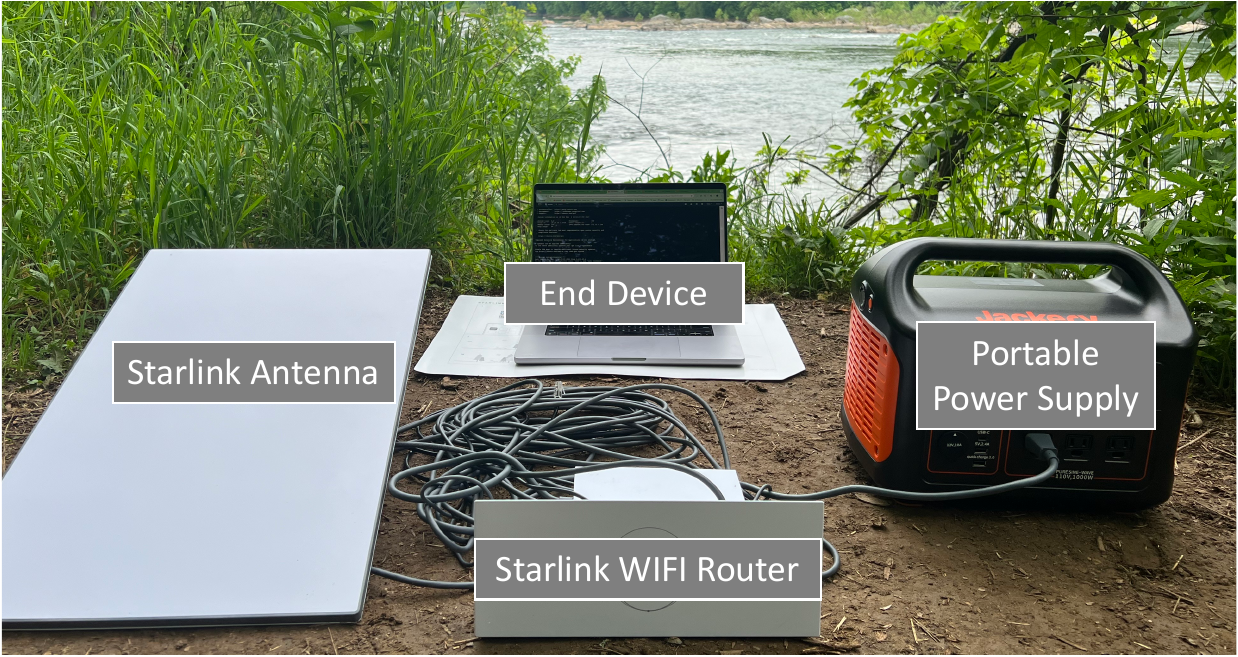}
\caption{\sysname Communication Test-Bed}
\label{fig: testbed}
\end{figure}

\begin{figure*}[t]
\centering
\includegraphics[width=0.95\linewidth]{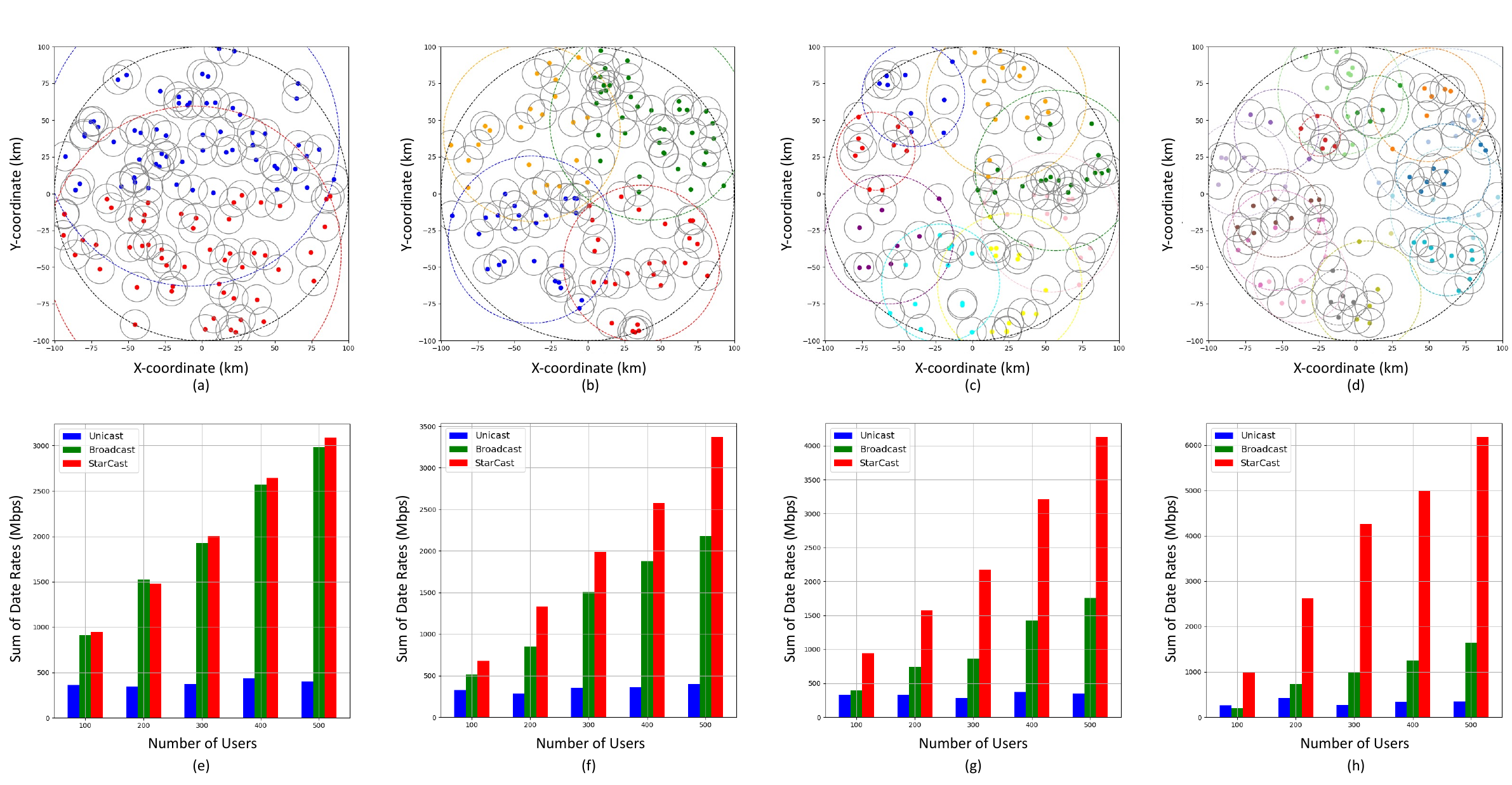}
\caption{
User Grouping and Beam Coverage in a Satellite-Serviced Area: Visualizations with (a) 2 Groups of Users, (b) 4 Groups of Users, (c) 8 Groups of Users, and (d) 16 Groups of Users; 
Performance Comparison of Sum of Data Rates for Legitimate Users Across the Three Communication Methods for (e) 2 Groups of Users, (f) 4 Groups of Users, (g) 8 Groups of Users, and (h) 16 Groups of Users.
}
\label{fig: data_rate}
\vspace{-0.2cm}
\end{figure*}

\begin{figure*}[t]
\centering
\includegraphics[width=0.95\linewidth]{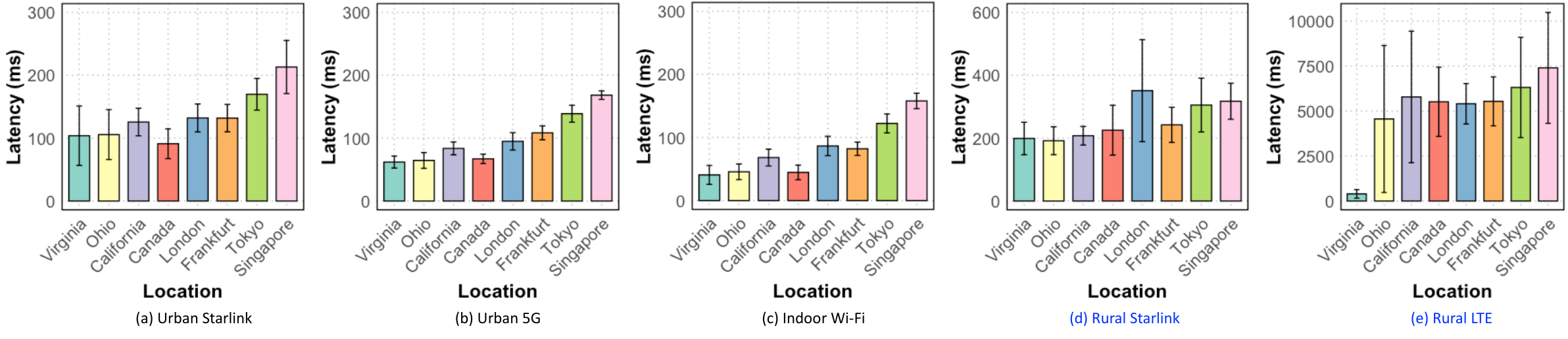}
\caption{Average ciphertext communication latency from eight AWS servers to the end device, with standard deviation samples indicated.}
\label{fig: latency}
\vspace{-0.2cm}
\end{figure*}

\begin{figure*}[t]
\centering
\includegraphics[width=0.95\linewidth]{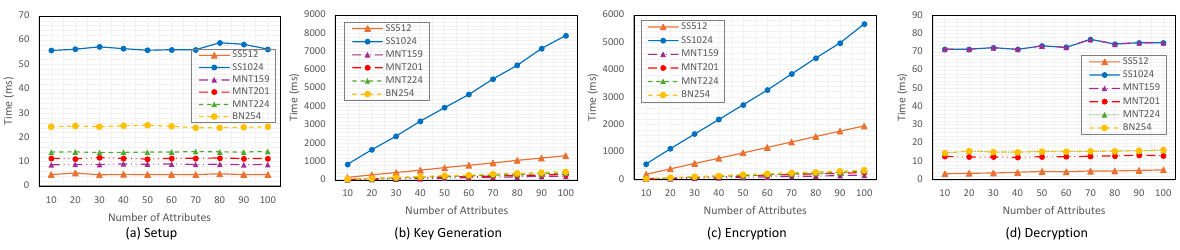}
\caption{Computation times for \sysname by (a) setup, (b) key generation, (c) encryption, and (d) decryption processes as the number of attributes increases.}
\label{fig: computation}
\end{figure*}

In comparing the sum of legitimate users' data rates of secure unicast, broadcast, and \sysname satellite transmission frameworks, we simulate the scheme with users distributed by groups in the coverage of the satellite, and the channel between satellite and user is modeled as the Rician channel. The system parameters for a typical LEO satellite are summarized in Tab.~\ref{tab: LEO_parameters}.

\begin{table}[h]
\centering
\caption{LEO Satellite Simulation Parameters}

\begin{tabular}{p{4.5cm}|p{2cm}}
\hline
\textbf{Parameters} & \textbf{Values} \\ \hline
The height of satellite $h$ & 550 km \\
The maximum beam coverage radius & 100 km \\   
Frequency Band & Ku (12 GHz) \\  
Bandwidth $B$ & 1 GHz \\  
Satellite power $P^t$ & 13 dBW \\  
User terminal antenna gain $G^r$ & 30 dBi \\  
The efficiency of the antenna $\eta$ & 0.7 \\  
Phased array antenna diameter $D$ & 0.5 m \\  
Power attenuation $\alpha$ & 1 \\  
AWGN power spectral density $N_0$ & -174 dBm/Hz \\ \hline

\end{tabular}
\label{tab: LEO_parameters}
\vspace{-0.2cm}
\end{table}

Fig.~\ref{fig: data_rate} (a)-(d) illustrate the beam coverage and user grouping scenarios for 2, 4, 8, and 16 groups of users, respectively, in a satellite-serviced area. Users are represented as colored dots corresponding to their groups, while the large black circle represents the broadcast beam coverage, the medium-sized circles denote the \sysname group communication beam coverage for each group, and the small gray circle indicates the unicast beam coverage area. As the number of user groups increases, the \sysname beam coverage becomes finer, improving beam directionality and targeting precision for legitimate groups.

In Fig.~\ref{fig: data_rate}(e)-(h), the simulated results compare the total data rates for legitimate users across unicast, broadcast, and \sysname frameworks for 2, 4, 8, and 16 groups of users, respectively. Each individual figure depicts the total data rates for legitimate users across different communication frameworks as the number of users increases. The unicast approach, while capable of providing the highest individual data rates due to its narrow and highly focused beams (resulting in a high antenna gain $G_t^i$), often yields the lowest sum of data rates for a group of legitimate users $C_m$. This inefficiency arises from the sequential time-sharing required to serve each user, leading to redundant transmissions of the encrypted content. Conversely, the broadcast framework delivers content to all users simultaneously by utilizing a wide beam with a large beamwidth $\theta_m$. However, this method results in a lower $C_m$ due to the reduced antenna gain $G_t^m$, as it must cover a broader area, which includes unintended users.

\sysname demonstrates its scalability and efficiency by adjusting the beamwidth \(\theta_m\) to optimally target specific user groups. This method enables simultaneous delivery of encrypted data to all legitimate users within each group. As the number of groups increases, \sysname beams become finer, effectively enhancing signal strength and reducing interference. Consequently, the \sysname framework achieves the highest \(C_m\) among the three communication methods as the number of user groups increases. It also demonstrates the efficiency with the number of groups of users increasing. As shown in Fig.~\ref{fig: data_rate}(e)-(h), the \sysname framework becomes increasingly efficient as the number of groups grows. While providing modest gains for 2 groups (Fig.~\ref{fig: data_rate}(e)), it achieves significant performance improvements for 4, 8, and 16 groups (Fig.~\ref{fig: data_rate}(f)-(h)), consistently outperforming unicast and broadcast in total data rates. This demonstrates the scalability and efficiency of \sysname in handling higher user densities within the group.

\subsection{Prototyping and Evaluation of \sysname Latency Using Starlink LEO Satellite Communication}

Currently, Starlink satellites with DTC capability support only emergency text services and are expected to extend to voice and IoT after 2025~\cite{starlinkdtc}. We prototyped \sysname on the Starlink Gen 3 Standard Kit and its LEO satellite network to measure latency (see Fig.~\ref{fig: testbed}). A MacBook Pro (Apple M1 Pro, 16 GB) served as the end device, connected to the Starlink antenna via WiFi. We also tested eight AWS virtual machines (free-tier EC2 instances, Ubuntu 22.04) located in Virginia, Ohio, California, Canada, London, Frankfurt, Tokyo, and Singapore to assess end-to-end ciphertext latency. Ciphertext was transmitted from these servers to the end device through the Starlink link. To study environmental effects, we deployed the antenna in both urban and rural areas. For comparison, we measured performance against T-Mobile 5G/LTE and indoor WiFi using the same AWS servers.

To establish a UDP broadcast session from an AWS server to a local device, we begin by creating a UDP socket on the AWS server using the \texttt{socket.socket(socket.AF\_INET, socket.SOCK\_DGRAM)} command. We enable the socket's broadcast capability by setting the \texttt{SO\_BROADCAST} option with \texttt{setsockopt(socket.SOL\_SOCKET, socket.SO\_BROADCAST, 1)}. After configuring the socket, we send the encrypted data to the broadcast address via the \texttt{sendto} method. On the local device side, we similarly set up a UDP socket with the \texttt{SO\_REUSEADDR} option enabled. This socket is then bound to listen on the same port as the server, and we use the \texttt{recvfrom} method to receive and process the broadcasted ciphertext. Although we implement the UDP broadcast on the Starlink network, the inherent limitation that Starlink does not physically support group communication in its current configuration means that our experimental results are based on an end-to-end session with a smaller beamwidth compared to a true group communication scenario. Therefore, our results are more representative of a situation where the legitimate group is geographically concentrated in a compact area.

\par \textbf{Latency Evaluation.} We first evaluate the communication latency from eight AWS servers to our end device located in an urban area, as shown in Fig.~\ref{fig: latency}(a)-(c). The latency increases as the distance from the server location to the client increases, ranging from Virginia to Singapore. Fig.~\ref{fig: latency}(a) presents the test results when the end device is connected to the Starlink antenna via a wireless signal. We observed that Starlink’s latency is, on average, slightly higher than that of the 5G network and indoor WiFi, as depicted in Fig.~\ref{fig: latency}(b)-(c). However, the variation in Starlink’s latency is significantly greater than that of the 5G network and indoor WiFi. We repeated the same experimental setup in a rural environment, where Starlink communications achieved latency below 400 ms, as shown in Fig.~\ref{fig: latency}(d). Notably, a latency spike occurs during communication from London, corresponding to a period when a connected satellite was moving away from the local antenna, according to the Starlink Coverage Map. In this rural location, 5G networks were unavailable, and the only terrestrial wireless network service was a weak T-Mobile LTE signal. Fig.~\ref{fig: latency}(e) shows that this unstable LTE signal exhibits significantly higher average latency and greater variation compared to other methods. In contrast, Starlink consistently provides reliable network service, even in rural areas.

\subsection{Computation Evaluation}

\par We implemented the \sysname schemes using Python 3.6 and the Charm 0.50 framework \cite{akinyele2013charm}. To assess the computation cost of \sysname, we conducted tests over elliptic curves of types SS512, SS1024, MNT159, MNT201, MNT224, and BN254, with the number of attributes ranging from 10 to 100. The experimental results are depicted in Fig.~\ref{fig: computation}. Fig.~\ref{fig: computation}(a) illustrates the time required for the setup algorithms of the CP-ABE schemes we implemented, with setup times consistently below 100 ms regardless of the number of attributes. Fig.~\ref{fig: computation}(b) and (c) show that the key generation and encryption times scale linearly with the number of attributes, corroborating our cryptographic design. The decryption time, in Fig.~\ref{fig: computation}(d) remains under 80 ms. This is because only a constant number of pairing operations are required. 

\subsection{Comparison of Encryption Efficiency Across Unicast, Broadcast, and \sysname Frameworks}

\textbf{Total Number of Ciphertexts.}  
In a unicast framework, where a satellite directs a narrow beam to individual users, delivering a message to a group $\mathcal{G}_m$ requires $|\mathcal{G}_m|$ encrypted messages, as each user receives a unique ciphertext with E2EE encryption~\cite{weiler2001secure, grosch2000framework}. In contrast, broadcast encryption requires only one ciphertext for all recipients, regardless of group size~\cite{barth2006privacy,fiat1994broadcast}. Similarly, \sysname transmits a single encrypted message, but only users meeting the access policy can decrypt it.

\textbf{Total Size of Ciphertexts.}  
In unicast, each user receives a unique ciphertext, making the total size $N$ times that of a single ciphertext with asymmetric encryption, where $N$ is the number of users. For example, with RSA-256, the total size is $N \times 2048$ bits. Broadcast encryption~\cite{barth2006privacy} scales linearly with group size, as the group key is encrypted with each member’s public key, resulting in a size of $N \times 2048$ bits plus overhead.  

The \sysname framework, leveraging attribute-based encryption, keeps ciphertext size dependent only on the number of attributes. As discussed in Sec.~\ref{ABE example}, an access policy requires only a small set of attributes (e.g., three in our case). \sysname's ciphertext comprises three elements from group $\mathbb{G}$ per attribute and three from group $\mathbb{H}$, ensuring efficient scaling. For an elliptic curve MNT224, each $\mathbb{G}$ element is 224 bits, while each $\mathbb{H}$ element is 672 bits. Comparatively, CGW~\cite{chen2015improved} requires two $\mathbb{G}$ and two $\mathbb{H}$ elements per attribute, leading to slightly larger ciphertexts, as $\mathbb{H}$ elements are three times the size of $\mathbb{G}$. Waters~\cite{waters2011ciphertext} and BSW~\cite{bethencourt2007ciphertext} further increase ciphertext size, requiring an attribute-dependent number of elements in $\mathbb{G}$ and an additional element in $\mathbb{H}$.

\textbf{Total Number of Encryption Keys.}  
In unicast, $N$ public keys are required, one per user. Similarly, broadcast encryption necessitates $N$ keys, as the group key is encrypted with each user's public key. In contrast, \sysname requires only one public key for encryption, embedding the access policy in the ciphertext to determine eligible decryption users.

\section{Conclusion}

In this paper, we present \sysname, a secure and efficient group encryption scheme for LEO satellite networks that addresses scalability, security, and spectrum efficiency. Leveraging CP-ABE, \sysname enables fine-grained access control, simplifies key management, and supports simultaneous transmissions to multiple users within a beam. This approach maximizes spectrum utilization while ensuring backward secrecy, revocability, and resistance to collusion. Our implementation and comparative evaluations against unicast and broadcast frameworks show that \sysname significantly improves the performance and efficiency of secure satellite communications. \sysname offers a scalable and reliable solution to meet growing connectivity demands in remote and underserved regions, laying the foundation for advanced IoT, disaster recovery, and military applications.

\section{Acknowledgment}

\par This work was supported in part by the Office of Naval Research under grants N00014-24-1-2730 and N00014-19-1-2621, the National Science Foundation under grants 2312447, 2247560, 2154929, 2332675, 2331936, and 2235232, and the Virginia Commonwealth Cyber Initiative (CCI).

\bibliographystyle{ieeetr}
\bibliography{IEEEabrv,ref}

\end{document}